\documentclass[aps,twocolumn,epsfig,rotate,showpacs]{revtex4}
\usepackage{epsfig}
\usepackage{graphicx}
\usepackage{amsmath}

\newcommand{\beq}{\begin{eqnarray}}
\newcommand{\eeq}{\end{eqnarray}}
\begin{document}

\title{Quantum Ratchet Accelerator without a Bichromatic Lattice Potential}
\author{Jiao Wang$^{1,2}$ and Jiangbin Gong$^{3,4}$}
\email{phygj@nus.edu.sg} \affiliation{$^{1}$Temasek Laboratories,
National University of Singapore, 117542, Singapore
\\$^{2}$Beijing-Hong Kong-Singapore Joint Center for Nonlinear and Complex Systems
(Singapore), National University of Singapore, 117542, Singapore
\\ $^{3}$Department of Physics and Center of Computational Science and Engineering,
National University of Singapore, 117542, Singapore
\\ $^{4}$NUS Graduate School for Integrative Sciences and Engineering, Singapore
 117597, Singapore}
\date{\today}

\begin{abstract}
In a quantum ratchet accelerator system, a linearly increasing
directed current can be dynamically generated without using a biased
field. Generic quantum ratchet acceleration with full classical
chaos [Gong and Brumer, Phys. Rev. Lett. 97, 240602 (2006)]
constitutes a new element of quantum chaos and an interesting
violation of a sum rule of classical ratchet transport.  Here we
propose a simple quantum ratchet accelerator model that can also
generate linearly increasing quantum current with full classical
chaos. This new model does not require a bichromatic lattice
potential. It is based on a variant of an on-resonance kicked-rotor
system, periodically kicked by two optical lattice potentials of the
same lattice constant, but with unequal amplitudes and a fixed phase
shift between them.  The dependence of the ratchet current
acceleration rate on the system parameters is studied in detail. The
cold-atom version of our new quantum ratchet accelerator model
should be realizable by introducing slight modifications to current
cold-atom experiments.

\end{abstract}

\pacs{05.45.Mt, 05.45.-a, 05.60.Gg, 32.80.Qk} \maketitle

\section{introduction}

A ratchet accelerator (RA) \cite{Gongpre04} can generate, without
using a biased field, directed transport in both the momentum and
coordinate space.  Specifically, certain spatio-temporal symmetries
in the Hamiltonian dynamics are broken and as a result a linearly
increasing directed current can be dynamically generated. Such a
property of RA systems is of considerable interest for understanding
(i) general properties of quantum and classical ratchet effects in
Hamiltonian systems
\cite{Gongpre04,schanzprl,tania,cheon,schanzpre,ratchetreview,denisov,lundh,kenfack},
(ii) quantum-classical correspondence in transport phenomena
\cite{schanzprl,tania,schanzpre}, and (iii) a number of interesting
topics in quantum chaos \cite{casatibook}.

Ongoing cold-atom studies of the well-known quantum kicked-rotor
(QKR) model \cite{casatibook} have motivated several RA studies
using QKR variants. In particular, Ref. \cite{lundh} showed that
accelerating quantum ratchet current can be realized by considering
a QKR variant, with the kicking period on the main resonance with
the recoil frequency of the cold atoms. Reference \cite{kenfack}
showed that a quantum RA can also be realized with QKR variants on
high-order quantum resonances. In both cases, spatio-temporal
symmetries in the dynamics are broken by using a  bichromatic
optical lattice (specifically, an optical super-lattice obtained by
superimposing two stand-waves with periods $\lambda/2$ and
$\lambda/4$).  However, though already achieved in some static cases
\cite{weitz}, experimentally realizing bichromatic and pulsed
optical lattices is somewhat demanding. Indeed, in two recent
cold-atom on-resonance-QKR experiments \cite{sadgrove,danaprl},
directed quantum transport is effectively demonstrated by use of a
single-period optical lattice only, with the price being that
special symmetry-breaking initial superposition states should be
prepared.

The ratchet transport in the above-mentioned on-resonance-QKR models
\cite{lundh,sadgrove,danaprl,kenfack,baowen} occurs only for
isolated values of the effective Planck constant (to be defined
below). By contrast, using variants of another paradigm of quantum
chaos, namely, the kicked-Harper model \cite{KHM}, Gong and Brumer
\cite{Gongprl06} proposed a quantum RA model that works for an
arbitrary value of the effective Planck constant. In this sense, the
ratchet transport in this new model \cite{Gongprl06} is {\it
generic}. Furthermore,  this generic RA model works even when the
underlying classical dynamics is fully chaotic, a situation where
classical ratchet transport necessarily vanishes according to a
classical ``sum rule" \cite{schanzpre,schanzprl}. Hence the work in
Ref. \cite{Gongprl06} represents an interesting and generic quantum
violation of a classical theorem.

The detailed aspects of the above-mentioned quantum violation of the
classical sum rule are yet to be explored. Along this direction, a
cold-atom realization of a generic quantum RA model would be of
great interest. Nevertheless, such experiments were thought to be
challenging because the model proposed in Ref. \cite{Gongprl06} also
employed a flashing bichromatic optical lattice and it was unclear
how a kicked-Harper-like model can be realized in a cold-atom laboratory.

Thanks to our recent finding \cite{jiaopra08,jiaojmo} that exposed a
direct connection between QKR and a class of kicked-Harper-like
models, here we are able to propose a quantum RA model that (i)
contains all the important ingredients as the model proposed in Ref.
\cite{Gongprl06}, (ii) does not require a bichromatic lattice
potential,  and (iii) is realizable by slightly modifying existing
cold-atom experiments of QKR dynamics.  Indeed, this new RA model
only requires an on-resonance variant of QKR, kicked by two optical
lattice potentials of the same lattice constant, but with unequal
kicking amplitudes and a fixed phase shift between them. In addition
to offering a simpler quantum RA model that is of theoretical
interest, it is hoped that our results below will motivate cold-atom
experimental studies in the near future.

This paper is organized as follows. In Sec. II we show how a wide
class of twisted kicked Harper models can be realized by using an
on-resonance ``double-kicked" rotor model.  General discussions in
Sec. II directly lead to an atom-optics proposal for realizing the
RA model proposed in Ref. \cite{Gongprl06}. In Sec. III we simplify
the cold-atom RA realization in Sec. II, resulting in a new RA model
that does not need a bichromatic optical lattice potential.  We then
present and discuss detailed numerical results of our new RA model,
with an emphasis placed on the dependence of the current
acceleration rate on the system parameters.  In Sec. IV we briefly
discuss one extension of this study. Section V concludes this work.

\section{Cold-Atom Realizations of A Wide Class of Twisted Kicked Harper
Models}

Our starting point is the so-called double kicked-rotor model (DKRM)
\cite{DKR1,DKR2,DKR3} that has been experimentally realized.  We use
scaled and dimensionless variables throughout. The DKRM Hamiltonian
is then given by
\begin{eqnarray}
H_{\text{DKRM}} &=& \frac{p^{2}}{2}+ K V_{K}(q)\sum_{n}\delta (t-nT) \nonumber \\
&+& LV_{L}(q)\sum_n \delta(t-nT-\eta),
\end{eqnarray}
where $q$ ($\in [0,2\pi)$) and $p$ are conjugate coordinate and
momentum operators, $T$ is the period for both kicking sequences,
$\eta$ is the time delay between the two kicking sequences, $K$ and
$L$ characterize the amplitudes of the kicking fields, $ V_{K}(q)$
and $ V_{L}(q)$ are periodic functions of $q$ with the period
$2\pi$.  The associated quantum map $U_{\text{DKRM}}$ for a period
from $nT+0^{-}$ to $(n+1)T+{0}^{-}$ is given by
\begin{eqnarray}
U_{\text{DKRM}}=e^{-i(T-\eta)\frac{p^2}{2\hbar}}e^{-i\frac{L}{\hbar}V_{L}(q)}
e^{-i\eta\frac{p^2}{2\hbar}}e^{-i\frac{K}{\hbar}V_{K}(q)},
\label{QM}
\end{eqnarray}
where $\hbar$ represents an effective and dimensionless Planck
constant for the DKRM system (hence $p=-i\hbar\partial/\partial
q$).

Theoretically, we shall first focus on an ideal situation where
cold atoms are injected with exactly zero quasi-momentum
\cite{quasimom}. With that simplification we may consider only a
Hilbert space satisfying the periodic boundary condition
associated with $q\rightarrow q+2\pi$. The quantum resonance
condition $T\hbar=4\pi$ then leads to
\begin{eqnarray} e^{-iT\frac{p^2}{2\hbar}}=1,
\end{eqnarray} reducing
$U_{\text{DKRM}}$ to $U_{\text{DKRM}}^{\text{r}}$,
\begin{eqnarray}
{U}^{\text{r}}_{\text{DKRM}}&=&
e^{i\eta\frac{p^2}{2\hbar}}e^{-i\frac{L}{\hbar}V_{L}(q)}
e^{-i\eta\frac{p^2}{2\hbar}}e^{-i\frac{{K}}{\hbar}V_{K}(q)}
\nonumber \\
&=&
e^{i\frac{\tilde{p}^2}{2\tilde{\hbar}}}e^{-i\frac{\tilde{L}}{\tilde{\hbar}}V_{L}(q)}
e^{-i\frac{\tilde{p}^2}{2\tilde{\hbar}}}e^{-i\frac{\tilde{K}}{\tilde{\hbar}}V_{K}(q)},
 \label{Ur}
\end{eqnarray}
where we have defined the rescaled momentum
\begin{eqnarray}
\tilde{p}\equiv \eta p
\end{eqnarray} and the rescaled kicking
amplitudes \begin{eqnarray}
 \tilde{K} &\equiv& \eta K, \label{Kvalue} \\
\tilde{L}&\equiv &\eta L.\label{Lvalue} \end{eqnarray}
 Due to the above momentum rescaling, the effective Planck constant now becomes
\begin{eqnarray}
\tilde{\hbar}\equiv \eta \hbar.
\label{etahbar}
 \end{eqnarray}

Equations (\ref{Kvalue},\ref{Lvalue},\ref{etahbar}) show that the
rescaled dimensionless system parameters $\tilde{K}$, $\tilde{L}$,
and $\tilde{\hbar}$ can be easily tuned by adjusting the time delay
between the two kicking sequences.  Based on a previous DKRM
experiment \cite{DKR1}, we estimate that in experiments the kicking
amplitudes $\tilde{K}$ and $\tilde{L}$ can vary in the range of
$0.1-100$, and the effective Planck constant $\tilde{\hbar}$ can at
least vary in the range of $0.05-20$.  Our computational studies in
the next section will be based on these two ranges.

To gain insights into the quantum-resonance-reduced quantum map in
Eq. (\ref{Ur}), let us first re-interpreted it as follows.  Reading
the four factors in Eq. (\ref{Ur}) from right to left, one sees that
within each period $T$, in effect the system is first subject to one
kick, followed by a free evolution of duration unity; then the
system is kicked a second time, followed by a second free evolution
of the same duration, but now with the free Hamiltonian given by
\begin{eqnarray}
H_{\text{free}}=-\tilde{p}^2/2.
\end{eqnarray}
Such an effective Hamiltonian with a negative kinetic energy term
was first considered in Ref. \cite{gongpre07}. With this
interpretation, one may define an ``$\eta$-classical" limit of this
quantum map, i.e., the $\tilde{\hbar}\equiv \eta \hbar\rightarrow 0$
limit with fixed $\tilde{K}$ and $\tilde{L}$. This terminology is
inspired by the so-called ``$\epsilon$-classical" limit in early
studies of QKR models in the presence of gravity \cite{Fishmancla}.
Let ${q}^{c}$ and $\tilde{p}^{c}$ be the counterparts of $q$ and
$\tilde{p}$ in this ``$\eta$-classical" limit,  with their values
right before $t=nT$ denoted by ${q}^{c}_n$ and $\tilde{p}^{c}_n$.
Further defining
\begin{eqnarray}
{\cal P}^{c}\equiv {q}^{c} + \tilde{p}^{c}, \end{eqnarray} one
easily finds the classical map associated with the
``$\eta$-classical" limit,
\begin{eqnarray}
{\cal P}_{n+1}^{c}& =& {\cal P}_{n}^{c} -
\tilde{K}\frac{dV_{K}({q}^{c}_n)}{d {q}^{c}_n} \\
{q}^c_{n+1}& =&  {q}^c_n + \tilde{L}\frac{dV_{L}({\cal
P}^{c}_{n+1})}{d{\cal P}^{c}_{n+1}}. \label{CKH}
\end{eqnarray}
In terms of the canonical pair $q^{c}$ and ${\cal P}^{c}$, the
classical Hamiltonian $H^{\text{c}}_{\eta}$ that generates this
``$\eta$-classical" map is then given by
\begin{eqnarray}
{H}^{\text{c}}_{\eta}=\tilde{L}V_{L}({\cal P}^c)+
\tilde{K}V_{K}(q^c)\sum_n\delta(t-n). \label{CHamiltonian}
\end{eqnarray}

Consider now the simplest choice for the kicking potentials, i.e.,
$V_{K}(q)=V_{L}(q)=\cos(q)$.  Such a choice under the restriction
$\tilde{K}=\tilde{L}$ was adopted by the original experiment
\cite{DKR1} and previous theoretical studies of off-resonance DKRM
\cite{DKR2, DKR3}. Substituting $V_{K}(q)=V_{L}(q)=\cos(q)$ into
Eq. (\ref{CHamiltonian}), the resulting ``$\eta$-classical"
Hamiltonian becomes precisely the classical kicked Harper model in
terms of ${\cal P}^{c}$ and $q^c$ \cite{jiaopra08}. Returning to
the old representation $(q^c, \tilde{p}^c)$, the obtained kicked
Harper Hamiltonian becomes
\begin{eqnarray}
{H}^{\text{c}}_{\text{TKH}}=\tilde{L}\cos(\tilde{p}^c+q^{c})+
\tilde{K}\cos(q^c)\sum_n\delta(t-n).
\end{eqnarray}
Comparing the Hamiltonian ${H}^{\text{c}}_{\text{TKH}}$ with the
standard kicked-Harper Hamiltonian as a function of $q^{c}$ and
$\tilde{p}^c$, i.e.,
\begin{eqnarray} {H}^{\text{c}}_{\text{KH}}=\tilde{L}\cos(\tilde{p}^c)+
\tilde{K}\cos(q^c)\sum_n\delta(t-n), \label{sckh} \end{eqnarray} we
can regard ${H}^{\text{c}}_{\text{TKH}}$ as a ``twisted" version of
the standard kicked-Harper model ${H}^{\text{c}}_{\text{KH}}$. With
this in mind, the quantum map in Eq. (\ref{Ur}) in the case of
$V_{K}(q)=V_{L}(q)=\cos(q)$ can be regarded as a quantized version
of the twisted kicked Harper model.

We now apply this on-resonance DKRM strategy to realize a twisted
version of the quantum RA model proposed in Ref. \cite{Gongprl06}.
This RA model involves the classical Hamiltonian
\begin{eqnarray}
H^{\text{c}}_{\text{BKH}}&=&\tilde{L}\cos(\tilde{p}^{c}) \nonumber \\
&&
+\tilde{K}\left[\cos(q^{c}+\phi_1)+\sin(2q^{c}+\phi_2)\right]\sum_n\delta(t-n).
\nonumber \\
\ \
\end{eqnarray}
If we now consider the following scenario:
\begin{eqnarray}
&& \ \ \text{Scenario\ I:} \nonumber \\
&&  V_{L}(q)=\cos(q); \\
&&  V_{K}(q)= [\cos(q+\phi_1)+\sin(2q+\phi_2)],
\end{eqnarray} then a twisted version of
$H^{\text{c}}_{\text{BKH}}$ (i.e., in terms of ${\cal P}^{c}$ and
$q^{c}$ rather than $\tilde{p}^{c}$ and $q^{c}$) naturally emerges
from Eq. (\ref{CHamiltonian}).  Clearly then, at least for a
twisted version, a cold-atom quantum version of the bichromatic
generalized kicked Harper model $H^{\text{c}}_{\text{BKH}}$ is
realizable, provided that a kicking bichromatic lattice potential
such as $[\cos(q+\phi_1)+\sin(2q+\phi_2)]\sum_n\delta(t-n)$ can be
realized. In the next section a simpler realization of quantum RA
is obtained.

Before ending this section we make one important remark.  In the
standard kicked Harper model ${H}^{\text{c}}_{\text{KH}}$ in Eq.
(\ref{sckh}), the momentum variable is an {\it abstract} canonical
variable. This becomes obvious if we consider the canonical
equations of motion,  yielding that the moving speed in the
coordinate space is not proportional to the momentum. As such,  it
is unclear whether the momentum variable in the kicked Harper model
can be directly related to the mechanical momentum of a moving
particle. Dana managed to connect this abstract momentum variable
with the mechanical momentum of a charged particle kicked by a
special sequence of magnetic fields \cite{dana}. Here, through the
cold-atom realization of a wide class of twisted kicked Harper
models, we are linking the momentum variable in the kicked Harper
model with the mechanical momentum of cold atoms.  Only through such
connections can the expectation value of the momentum be interpreted
as a current of moving particles.

\section{Ratchet Accelerator without a Bichromatic optical
Lattice}

Our discussions in the previous section make it clear that, in
realizing a wide class of kicked-Harper-like models with
on-resonance DKRM,  the following
 canonical transformation or twist is necessarily involved:
\begin{eqnarray}
(q^c, \tilde{p}^{c})\rightarrow (q^{c}, {\cal P}^{c}).
\end{eqnarray}
Due to this phase space twist,  the resultant systems should assume
different symmetry properties than those analyzed in terms of
$q^{c}$ and ${\tilde p}^{c}$.  Hence, the symmetry-breaking
considerations in Ref. \cite{Gongprl06} no longer apply to twisted
kicked-Harper-like models. As a result,  the use of a bichromatic
optical lattice as in Ref. \cite{Gongprl06} may not be the simplest
approach for symmetry-breaking.  It is this recognition that
motivated us to seek a realization of a quantum RA without using a
bichromatic lattice potential.  This attempt is also consistent with
a recent study of ratchet transport (in coordinate space only) using
an off-resonance DKRM involving two optical lattices of the same
lattice constant \cite{carlos}.

Specifically, here we shall demonstrate that an on-resonance DKRM
with the following alternative scenario,
\begin{eqnarray}
&&\ \ \text{Scenario\ II:} \nonumber \\
&& V_{K}(q)=\cos(q);  \\
&& V_{L}(q)=\cos(q+\phi)
\end{eqnarray} can already
give rise to a simple quantum RA model if $K\ne L$.  In addition to
the on-resonance condition, this scenario only needs to introduce
two small modifications to a previous DKRM experiment \cite{DKR1}.
First, the two kicking sequences of optical lattice potentials
should have different amplitudes. Second, there should be a fixed
phase shift $\phi$  between these two optical lattice potentials.

Using Eq. (\ref{Ur}), scenario II described above gives the
following quantum map,
\begin{eqnarray}
{U}_{\text{RA}}=
e^{i\frac{\tilde{p}^2}{2\tilde{\hbar}}}e^{-i\frac{\tilde{L}}{\tilde{\hbar}}\cos(q+\phi)}
e^{-i\frac{\tilde{p}^2}{2\tilde{\hbar}}}e^{-i\frac{\tilde{K}}{\tilde{\hbar}}\cos(q)}.
 \label{URA}
 \end{eqnarray}
Using Eq. (\ref{CHamiltonian}), one then obtains the
$\eta$-classical Hamiltonian of this quantum map,
\begin{eqnarray}
H^{\text{c}}_{\text{RA}}=\tilde{L}\cos(q^{c}+\tilde{p}^{c}+\phi)+
\tilde{K}\cos(q^c)\sum_n\delta(t-n). \label{RACHamiltonian}
\end{eqnarray}

Let $|n\rangle$ be the eigenstates of the momentum operator $\tilde
p$, with an eigenvalue $n \tilde\hbar$ for a Hilbert space with the
periodic boundary condition. In the following we shall focus on the
RA dynamics for the initial state $|0\rangle$. The classical analog
of this initial state is a classical ensemble with $\tilde{p}^{c}=0$
and a random uniform distribution of $q^{c}$. Such an initial state
is a trivial state because it is symmetric upon time-reversal
operations or space-reflection operations. With this choice of the
initial state, any induced current afterwards must be due to some
broken spatio-temporal symmetries in the ensuing dynamics.   It is
also worth noting that the system described by Eq. (\ref{URA}) is
invariant under the transformations $q\to 2\pi-q$, $\tilde p\to
-\tilde p$ and $\phi\to -\phi$. As a result the current should
undergo a sign change under $\phi\to -\phi$, leading to the
expectation that the generation of a ratchet current is forbidden
for $\phi=0$ or $\phi=\pi$.  For this reason we focus on other
values of $\phi$.

\subsection{Examples of Accelerating Ratchet Current}

Consider first a few computational examples depicted in Fig. 1.
There, the time dependence of the quantum ratchet current, i.e., the
expectation value of the scaled momentum, $\langle
\tilde{p}\rangle$, is shown for some particular values of
$\tilde{K}$, $\tilde{L}$, and $\phi$.  The initial state is
$|0\rangle$ and the time propagation due to the quantum map
$U_{\text{RA}}$ is carried out by standard
fast-Fourier-transformation techniques. The cases shown in Fig. 1
display spectacular linear acceleration of the ratchet current at a
significant rate.  In order to have a comparison with the underlying
$\eta$-classical limit, we have also calculated the
ensemble-averaged classical momentum, $\langle {\tilde p}^c
\rangle$. The classical calculations are based on the
$\eta$-classical map given by Eq. (\ref{RACHamiltonian}), using an
ensemble of $10^6$ particles initially distributed along ${\tilde
p}^c=0$ randomly and uniformly. As is seen from Fig. 1a, the
$\eta$-classical currents can also increase linearly, with a slope
smaller than their quantum counterparts.  The result in Fig. 1b is
even more interesting. There the classical current remains
indistinguishable from zero at all times, but the quantum
acceleration is substantial. Note also that these
``$\eta$-classical" results have nothing to do with the true
classical limit of a DKRM, because here the DKRM is always on the
main quantum resonance. Indeed, the $\eta$-classical Hamiltonian is
given by Eq. (\ref{RACHamiltonian}), whereas the true classical
Hamiltonian of the DKRM should take exactly the same form as Eq.
(1).

\begin{figure}
\begin{center}
\epsfig{file=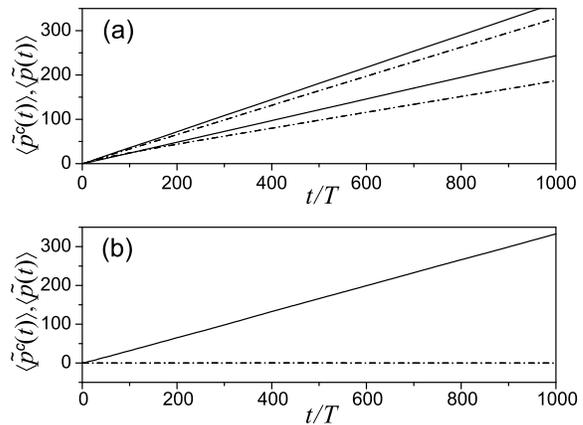,width=8.4cm}\caption{Time dependence of the
ratchet current for both the quantum ratchet accelerator model of
Eq. (\ref{URA}) (solid lines) and its $\eta$-classical limit
(dash-dotted lines) described by Eq. (\ref{RACHamiltonian}). For
panel (a), system parameters are $\tilde \hbar=1$, $\tilde K=3$,
$\tilde L=1$, and $\phi=\pi/2$ (upper two curves), $\pi/3$ (bottom
two curves). For panel (b), system parameters are $\tilde \hbar=1$,
$\tilde K=4$, $\tilde L=2$, and $\phi=\pi/2$. The classical current
in panel (b) remains indistinguishable from zero at all times
because the system is in the full chaos regime (see Fig. 3d).}
\label{fig1} \end{center}
\end{figure}

We have also checked that if we choose $\phi=0,\pi$ instead, then
both the classical and quantum acceleration seen in the examples in
Fig. 1a do vanish. This confirms our previous discussion on a
symmetry property of our new RA model.

\subsection{Dependence of Acceleration Rate on $\tilde{K}$ and $\tilde{L}$}
To further explore the dynamical aspects of our new RA model, we
have carried out detailed studies of how the ratchet acceleration
rate depends on the system parameters. The computational examples
shown in Fig. 1 motivate us to define the quantum current accelerate
rate as follows:
\begin{eqnarray}
R_q \equiv d \langle \tilde  p  (t)\rangle/dt.
\end{eqnarray}
For the sake of comparison we also define the $\eta$-classical
current acceleration rate as \begin{eqnarray} R_c=d\langle \tilde
p^c (t)\rangle /dt.
\end{eqnarray}
Computationally, these rates are determined as the average linear
increase rate over the time range $1000\le t/T<2000$.  Once the
linear acceleration rates are obtained, we then check, in many
cases, to see if the dynamics over a much longer time scale still
accelerates the current with the same rate. Most often this is
indeed the case, but some negative cases due to transient effects
will be mentioned below.

\begin{figure}
\vspace{0.1cm}\epsfig{file=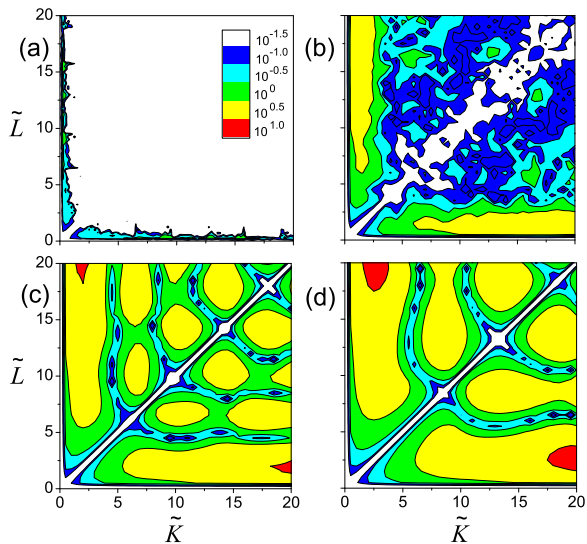,width=10.5cm} \caption{(Color
online) (a): Absolute values of the classical current acceleration
rate, denoted $\left|R_{c}\right|$,  as a function of $\tilde{K}$
and $\tilde{L}$, for the RA model in the $\eta$-classical limit,
Eq. (\ref{RACHamiltonian}). (b)-(d): Absolute values of the
quantum current acceleration rate, denoted $\left|R_{q}\right|$,
as a function of $\tilde{K}$ and $\tilde{L}$, for the RA model in
Eq. (\ref{URA}), with (b) $\tilde \hbar=1$, (c) $\tilde
\hbar=2\pi/3$, and (d) $\tilde \hbar=\pi$.  In all the cases we
set $\phi=\pi/2$. The contour scale $10^{-1.5}$ is for values less
than $10^{-1.5}$, the contour scale $10^{-1.0}$ is for values
between $10^{-1.5}$ and $10^{-1.0}$, and so on. The contour scales
used in panel (a) apply to other panels as well.}\label{fig2}
\end{figure}

Figure 2 shows the contour plots of $\left|R_q\right|$ and
$\left|R_c\right|$ thus obtained as a function of $\tilde K$ and
$\tilde L$.   A number of interesting features can be observed from
Fig. 2. First, significant classical ratchet acceleration (see Fig.
2a) exists only for those parameter regimes close to the $\tilde{K}$
axis or the $\tilde{L}$ axis.  That is, at least one of the two
values of $\tilde K$ and $\tilde L$ should be small for a
considerable classical ratchet acceleration to emerge. But even that
condition does not suffice. It is also clear from Fig. 2a that the
regime of $\tilde{K}\sim \tilde{L}$ should be excluded in order to
have an appreciable $\left|R_c\right|$. The overall result is that
in the parameter space defined by $\tilde K$ and $\tilde L$, only a
very small portion can yield considerable ratchet acceleration in
the $\eta$-classical limit. Second, the quantum results shown in
Fig. 2b-2d display many interest patterns. These patterns are absent
in the classical case shown in Fig. 2a, and they vary strongly if we
change $\tilde{\hbar}/\pi$ from an irrational value to a rational
value. It can also be seen from Fig. 2b-2d that appreciable quantum
ratchet acceleration occurs in a much larger parameter regime, often
with $\left|R_q\right|>\left|R_c\right|$. Third, the quantum results
share one feature with the classical result. That is, along the
direction of $\tilde{K}=\tilde{L}$, $|R_q|$ is also seen to be small
(typically much smaller than $10^{-1.5}$).

Some exceptions seem to be captured by Fig. 2b, where
$\left|R_q\right|$ can become larger than $10^{-1.5}$ along the
direction $\tilde{K}=\tilde{L}$.  However, upon a careful
investigation,  we find that these exceptions are mainly caused by
the particular way we numerically determine $R_q$. Indeed, if we
follow the dynamics much longer (e.g. $10^4-10^6$ kicks), then the
ratchet current tends to saturate for these exceptional cases, in
contrast to the unbounded linear acceleration observed in other
cases with $\tilde{K}\ne \tilde{L}$. Detailed investigations of such
transient effects in the ratchet acceleration are beyond the scope
this work. The exact boundary between bounded and unbounded quantum
current acceleration can be an interesting and challenging
mathematical problem.

\begin{figure}
\epsfig{file=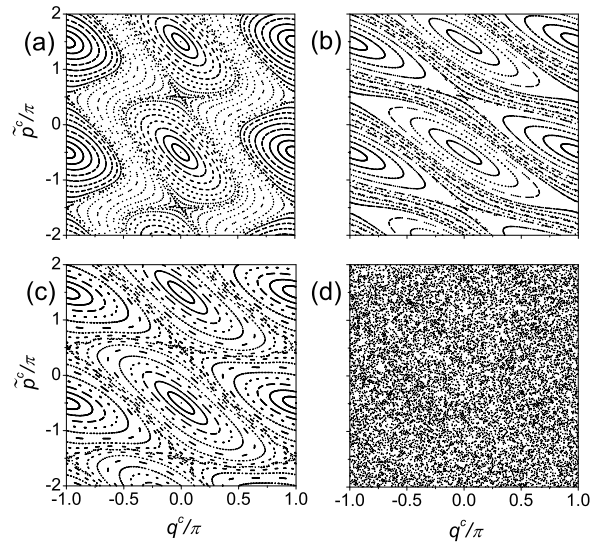,width=10.7cm}\vspace{0.0cm} \caption{The
phase space portrait of the $\eta$-classical limit described by Eq.
(\ref{RACHamiltonian}), for $2\tilde L=\tilde K=1$ in panel (a),
$2\tilde K=\tilde L=1$ in panel (b), $\tilde K=\tilde L=1$ in panel
(c), and $2\tilde L=\tilde K=4$ in panel (d). For all the panels  we
choose $\phi=\pi/2$. In cases (a) and (b) phase space invariant
curves extended in the momentum space can be clearly seen. In case
(c) there is a web of separatrix structures and phase space
invariant curves along the momentum space no longer exist. Case (d)
represents a fully chaotic phase space if both $\tilde{K}$ and
$\tilde{L}$ are sufficiently large. }\label{fig3}
\end{figure}

To shed more light on the results in Fig. 2, let us examine in
Fig. 3 the phase space structures of the $\eta$-classical limit of
our RA model. The phase space structure of the entire phase space
is just an infinite repetition (in both $\tilde{q}^c$ and
$\tilde{p}^c$) of what is shown in Fig. 3. If $\tilde K\ne \tilde
L$ and either $\tilde{K}$ or $\tilde{L}$ is sufficiently small,
then we always find phase space invariant curves extended in
momentum. Trajectories moving along these invariant curves will
display ballistic-like diffusion.  Such a phase space feature
differs from that of the standard kicked Harper model. In the
latter case the phase space invariant curves can lie parallel to
the $q^{c}$-axis if $\tilde{K}\ll\tilde{L}$.  This difference is
expected, because the $\eta$-classical Hamiltonian in Eq.
(\ref{RACHamiltonian}) is a {\it twisted} version of the kicked
Harper model.

Taking into account that $q^c=\pm \pi$ are equivalent points in the
phase space, one can easily see that in both cases of Fig. 3a and
Fig. 3b, there exist two bundles of phase space invariant curves,
separated by a seperatrix structure associated with some unstable
fixed points.  Remarkably,  the moving directions of the
trajectories on the two bundles are opposite to each other. This
feature is also consistent with the classical sum rule
\cite{schanzprl}.  Based on these observations we are ready to
explain the origin of the classical accelerating ratchet current. In
particular, because the overlap of the $\tilde p^c=0$ line (the
initial classical ensemble) with the two bundles of ballistic
diffusion curves can be different, the effects of the two bundles of
invariant curves cannot cancel out against each other and hence a
net current develops. The current will increase linearly with time
due to the ballistic nature of the phase space invariant curves
extended in momentum. This understanding is found to be consistent
with an estimate using the intersection lengths between the $\tilde
p^c=0$ line and the two bundles of phase space invariant curves.
This also clarifies the role of the parameter $\phi$. As is evident
from the expression of the $\eta$-classical Hamiltonian in Eq.
(\ref{RACHamiltonian}), the net effect of the parameter $\phi$ is a
shift of the phase space structure along the $\tilde{p}^{c}$-axis.
So the parameter $\phi$ can be used to tune the unbalanced overlap
of the initial ensemble with the two bundles of phase space
invariant curves.

By contrast, if $\tilde K=\tilde L$ (Fig. 3c), then before full
chaos sets in,  a web of separatrices emerge in the phase space and
there are no longer phase space invariant curves extended in
momentum.  As the value of $\tilde K=\tilde L$ increases, the
chaotic layer associated with the web of separatrices becomes
thicker and thicker. During this regular-to-chaos transition no
phase space invariant curves extended in momentum will emerge. As a
result, so long as $\tilde K=\tilde L$, ballistic-like diffusion
cannot happen and a linear acceleration of the ratchet current
becomes impossible. This directly explains why the case of $\tilde
K=\tilde L$ is so special for the $\eta$-classical ratchet
acceleration. Figure 2 also shows the absence of significant quantum
ratchet acceleration in cases of $\tilde K=\tilde L$.  We believe
that this quantum result is also due to the special classical phase
space structure for $\tilde K=\tilde L$.

Let us finally discuss the fully chaotic cases. One typical example
is shown in Fig. 3d. Analogous behavior can be found for other
larger values of $\tilde{K}$ and $\tilde{L}$. Clearly, phase space
invariant curves are all broken in these fully chaotic cases. As a
result, classical ratchet acceleration necessarily vanishes, as
illustrated in Fig. 1b.  This rationalizes the main message from
Fig. 2a, i.e., appreciable classical ratchet acceleration exists
only for a small fraction of the parameter space of $\tilde{K}$ and
$\tilde{L}$.

Remarkably, in general full classical chaos does not forbid quantum
ratchet acceleration. As shown in Fig. 2b for a generic value of
$\tilde{\hbar}$ (i.e., irrational with $\pi$), large
$\left|R_q\right|$ can be found for relatively large $\tilde{K}$ and
$\tilde{L}$, even when the associated $\eta$-classical dynamics
becomes fully chaotic. A specific computational example has been
shown in Fig. 1b. This hence confirms a similar observation made in
Ref. \cite{Gongprl06} and constitutes another example of generic
quantum violation of the classical sum rule \cite{schanzprl}.  In
Ref. \cite{Gongprl06}, this violation was explained in terms of the
concentration of quantum amplitudes on the remanents of classical
cantori-like phase space structures extended in momentum. Here, as
$\tilde{K}$ and $\tilde{L}$ increase, the two bundles of phase space
invariant curves will also generate classical cantori-like
structures. It can then expected that the classical cantori-like
structures  should be extended in the momentum space for
$\tilde{K}<\tilde{L}$ as well as $\tilde{L}<\tilde{K}$.
 As such,  the condition for quantum ratchet acceleration to occur in our new model
 should differ from that in Ref. \cite{Gongprl06}. Specifically, it is
unnecessary to have a sufficiently large ratio
$\tilde{K}/\tilde{L}$: a sufficiently small ratio
$\tilde{K}/\tilde{L}$ should also do. Given this understanding, the
quantum ratchet acceleration here with full classical chaos is
expected to display a ``dual" symmetry in the $\tilde{K}-\tilde{L}$
space. This is indeed the case: the pattern in Fig. 2b is symmetric
along the line of $\tilde{K}=\tilde{L}$.  Note also that for
non-generic values of $\tilde{\hbar}/\pi$,  the quantum ratchet
acceleration rates as shown in Fig. 2c and Fig. 2d can be even
larger, despite the fully chaotic phase space in the underlying
$\eta$-classical limit.

\begin{figure}
\begin{center}
\epsfig{file=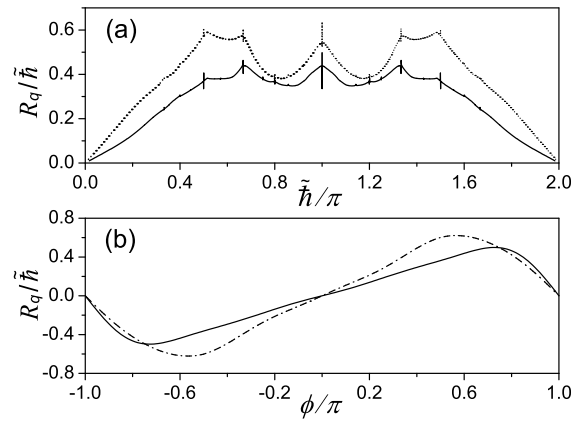,width=8.4cm}
\end{center}
\caption{(a) Dependence of the ratchet acceleration rate on the
effective Planck constant $\tilde \hbar$ for $\phi=\pi/2$ (dotted
line) and $\phi=1$ (solid line). (b) Dependence of the ratchet
acceleration rate on the system parameter $\phi$,  with $\tilde
\hbar=\pi/2$ (dash-dotted line) and $\tilde \hbar=1$ (solid line).
In all the cases here $3\tilde L=\tilde K=3\tilde\hbar$.}
\end{figure}

\subsection{Dependence of Acceleration Rate on $\tilde{\hbar}$ and $\phi$}

Focusing on the quantum case,  here we first examine the dependence
of $R_q$ on $\tilde \hbar$.  As already indicated by the drastic
differences between Fig. 2b, Fig. 2c, and Fig. 2d, one might wonder
if $R_q$ is too sensitive to $\tilde \hbar$ such that experimental
uncertainties in $\tilde \hbar$ may kill the ratchet acceleration
altogether.  To address this concern we show two typical
computational results in Fig. 4a for two values of $\phi$. In both
cases some sharp peaks of $R_q$ can be seen. These peaks are located
at those values of $\tilde \hbar$ that are rational multiples of
$\pi$. Nevertheless, the overall $\tilde\hbar$-dependence of $R_q$
does not show drastic oscillations.  They can be varying smoothly
over a considerably wide range of $\tilde{\hbar}$. This feature also
resembles what is found in Ref. \cite{Gongprl06}. In the regime of
very small $\tilde\hbar$, we have checked that $R_q$ does approach
the $\eta$-classical acceleration rate $R_c$, thus establishing the
expected quantum-classical correspondence.  It should be stressed
that the $\tilde{\hbar}$-dependence of $R_q$ shown here is a purely
quantum effect.  To have a theory accounting for this
$\tilde{\hbar}$-dependence  would be challenging but truly
fascinating.

A related question is whether or not the ratchet acceleration is
robust to variations in the parameter $\phi$ that characterizes
the phase lag between two optical lattices of the same lattice
constant. As demonstrated by the example shown in Fig. 4b, the
$\phi$-dependence of $R_q$ is smooth in the entire range of
$\phi$. This is somewhat expected: in the $\eta$-classical limit
the parameter $\phi$ only shifts the phase space structure along
the momentum axis. The conclusion is that small fluctuations in
$\phi$ or $\tilde\hbar$ should not be a big concern for
experimental studies.

\subsection{Effects of the Quasi-Momentum Spread in Cold-Atom Experiments}

In cold-atom experiments of kicked-rotor systems, the initial
state cannot be exactly the state with zero quasi-momentum, even
when the atoms are injected by a large-size Bose-Einstein
condensate.  Indeed, cold atoms are moving in real space, so their
matter-wave state does not need to satisfy the periodic boundary
condition inherent to a true kicked-rotor system.  To motivate
cold-atom experimental studies of our new RA model,  it becomes
necessary to examine the detrimental effects of the nonzero
quasi-momentum spread in the initial state.

Because the kicking optical lattice potentials are always periodic,
the quasi-momentum of the cold atoms, denoted $\beta$, is a
conserved quantity \cite{danaprl,quasimom}.  The dynamics emanated
from an initial state with a spread in $\beta$ can then be easily
simulated by considering each $\beta$-component separately.  To shed
more light on this issue let us return to the DKRM propagator
$U_{\text{DKRM}}$ in Eq. (\ref{QM}).  For each $\beta$ component,
one now has $\tilde p|n\rangle=(n+\beta)|n\rangle$.  This leads to
the consequence that $e^{-iT\frac{p^2}{2\hbar}}\ne 1$ under the
quantum resonance condition $T\hbar=4\pi$.   Nevertheless, for
$T\hbar=4\pi$ and for the potentials $V_{K}(q)$ and $V_{L}(q)$ used
in our RA model, it is enlightening to rewrite Eq. (\ref{QM}) as
\begin{eqnarray}
{U}_{\text{RA}}^{\beta}=e^{-i2\pi\frac{\tilde{p}^2}{\tilde{\hbar}^2}}
e^{i\frac{\tilde{p}^2}{2\tilde{\hbar}}}e^{-i\frac{\tilde{L}}{\tilde{\hbar}}\cos(q+\phi)}
e^{-i\frac{\tilde{p}^2}{2\tilde{\hbar}}}e^{-i\frac{\tilde{K}}{\tilde{\hbar}}\cos(q)}.
\label{URAB}
\end{eqnarray}
Except for the first factor,  this expression is completely parallel
to $U_{\text{RA}}$ in Eq. (\ref{URA}).  Because the first factor  is
no longer unity and changes with $\beta$, the first factor introduces
de-phasing when a distribution of $\beta$ values are averaged
over.  Therefore, it can be expected that the $\beta$ spread will
tend to saturate the accelerating ratchet current.

Below we assume a Gaussian distribution of $\beta$,  with the
variance denoted by $\Delta\beta$ and the mean value denoted by
$\overline{\beta}$.  Results for a typical case are shown in Fig. 5.
It is seen from Fig. 5a that a nonzero variance $\Delta\beta$ indeed
induces the saturation of the quantum ratchet current. The exact
saturation time increases, but slowly, with decreasing
$\Delta\beta$. For the parameters adopted in Fig. 5a, the typical
saturation time is around 20 kicking periods for $\Delta\beta \sim
0.002$ (scaled by $\tilde{\hbar}$), a characteristic value of the
$\beta$ spread reported in a recent experiment using Bose-Einstein
condensates \cite{ryu}. Figure 5b also shows an interesting
dependence of the ratchet current at $t=7T$ upon the mean
quasi-momentum $\overline{\beta}$ of the initial state. The result
for $\Delta\beta=0.01$ is seen to be almost the same as that for
$\Delta\beta=0.002$. This $\overline{\beta}$-dependence of the
ratchet current at early times might be of interest to experimental
studies as well.

In future experiments the quasi-momentum spread can be made smaller
than $\Delta\beta \sim 0.002$ \cite{ryu}. However, the result in
Fig. 5a indicates that even for $\Delta\beta \sim 0.0005$, the
saturation still sets in within a relatively short time scale.
Interestingly, a previous theoretical RA model using a bichromatic
optical lattice displays saturation at a similar time scale
\cite{kenfack}. This indicates that when faced with the detrimental
effects of the quasi-momentum spread,  the robustness of our new RA
model without a bichromatic optical lattice is similar to previous
models with bichromatic optical lattices.
\begin{figure}
\begin{center}
\epsfig{file=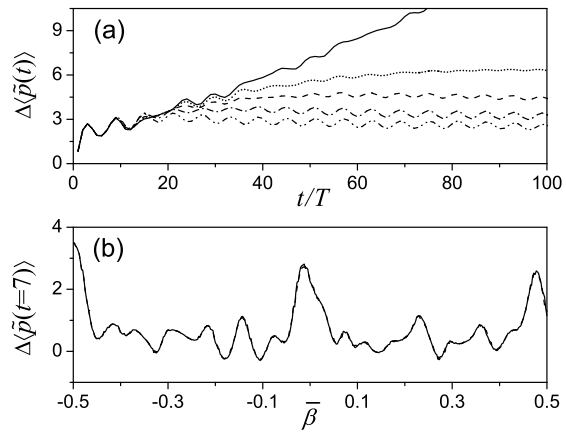,width=8.4cm}
\end{center}
\caption{Effects of the quasi-momentum spread on quantum ratchet
currents with $\tilde\hbar=1$, $2\tilde L=\tilde K=2$, $\phi=\pi/2$.
$\Delta \langle\tilde p(t)\rangle \equiv \langle \tilde p(t)\rangle
-\langle \tilde p(0)\rangle$. In panel (a), $\overline{\beta}=0$ and
the variance in $\beta$ is given by $\Delta \beta=0, 0.0005, 0.001,
0.002$ and $0.004$ (from above to bottom). In panel (b), $\Delta
\langle \tilde p(t)\rangle $ at $t=7$ with $\Delta \beta=0.002$
(dashed line) and $0.01$ (solid line) are shown as a function of the
mean quasi-momentum $\overline{\beta}$ of the initial state.}
\end{figure}

\section{Double-kicked rotor systems on high-order quantum
resonances}

So far we have studied the ratchet transport in the DKRM under the
main quantum resonance condition $T\hbar=4\pi$. To motivate both
theoretical and experimental studies in the future, in this short
section we briefly discuss an interesting extension of the current
study. The extension is about the dynamics of ratchet current
acceleration in DKRM on high-order quantum resonances. In such
extended cases, $T\hbar=4\pi\nu/\mu$, where $\nu$ and $\mu$ are two
incommensurate integers.  Under this high-order quantum resonance
condition we find that analogous ratchet acceleration can be
obtained as well, without using a bichromatic optical lattice. Thus,
ratchet current acceleration itself may provide a useful tool for
studies of high-order quantum resonances in DKRM.

\begin{figure}
\begin{center}
\vspace{0.0cm}\epsfig{file=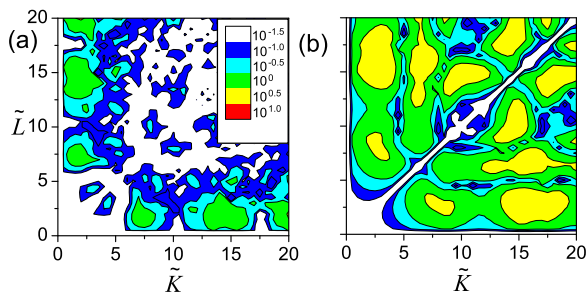,width=10.5cm}
\end{center}
\caption{ (Color online) Absolute values of the ratchet current
acceleration rate,  $|{R}_q|\equiv\left|d\langle \tilde{p}\rangle/dt
\right|$, for a double kicked rotor system at quantum anti-resonance
$T\hbar=2\pi$. The quantum propagator is given by Eq. (2). For the
sake of comparison, the same parameter rescaling as in the cases of
DKRM under the main quantum resonance is adopted. The rescaled
parameters are given by  $\tilde \hbar=1$ in panel (a), and $\tilde
\hbar=2\pi/3$ in panel (b). In both cases $\phi=\pi/2$. The meaning
of the contour scale is the same as in Fig. 2, and the contour
scales used in panel (a) apply to panel (b) as well.} \label{fig6}
\end{figure}

As an example in Fig. 6 we show $\left|R_q\right|$ under the quantum
anti-resonance condition $T\hbar=2\pi$, as a function of $\tilde{K}$
and $\tilde{L}$. The kicking potentials $V_{K}(q)$ and $V_{L}(q)$
are the same as those considered in Figs. 1-5. The case in Fig. 6a
represents cases with a generic value of $\tilde{\hbar}$, yielding
appreciable $\left|R_q\right|$ in some regimes. In this case the
detailed dependence of $\left|R_q\right|$ on $\tilde{K}$ and
$\tilde{L}$ is seen to be rather complicated. The case of
$\hbar=2\pi/3$ in Fig. 6b represents cases with nongeneric values of
$\tilde{\hbar}$. The associated ratchet acceleration effect is seen
to be larger over a wider regime. The dependence of
$\left|R_q\right|$ on $\tilde{K}$ and $\tilde{L}$ is also simpler
than that seen in Fig. 6a.  These results have no apparent
connections with classical ratchet transport, in the sense that for
high-order quantum resonances we can no longer define an
$\eta$-classical limit to guide our qualitative understandings.
However, as is evident from Fig. 6, even in these quantum
anti-resonance DKRM cases, a significant ratchet acceleration rate
also requires the condition $\tilde{K}\ne\tilde{L}$.  More studies
of this extension will be carried out in the near future.

\section{Concluding Remarks}

To conclude, we have proposed and studied a new quantum ratchet
accelerator model based on atom optics realizations of kicked-rotor
systems. Unlike all previous ratchet accelerator models, here we do
not need to use a bichromatic optical lattice potential. Based on
this advantage and the detailed computational studies presented
here, we believe that the cold-atom realization of our ratchet
accelerator model is within the reach of today's state-of-the-art
experiments \cite{sadgrove,danaprl}.  Indeed, the avenue of using
atom optics to experimentally study a whole class of
kicked-Harper-like models has been just opened up \cite{jiaopra08},
and the ratchet accelerator model proposed here seems to be a
wonderful starting point along this direction.

To have a linear acceleration of ratchet current in the
$\eta$-classical limit of our model, we have shown that the phase
space invariant curves extended in the momentum space are a
necessary condition.  Therefore, for kicking optical lattice
potentials $K\cos(q)$ and $L\cos(q+\phi)$,  an on-resonance
double-kicked rotor with equal kicking amplitudes $K$ and $L$ cannot
yield an unbounded and linearly increasing classical current.
Instead, we need unequal kicking amplitudes for accelerating and
unbounded classical current to occur. This interesting requirement
is also observed, but not fully explained, in the quantum dynamics.
Given unequal kicking amplitudes, the quantum ratchet acceleration
in our model can however persist for large kicking amplitudes, even
when the $\eta$-classical limit no longer has phase space invariant
curves extended in the momentum space. This purely quantum effect is
believed to be another example that remanents of classical phase
space structures can dramatically impact the quantum dynamics.
Considering these insights, we hope that our simple ratchet
accelerator model will also motivate future theoretical work to
better understand quantum transport and quantum-classical
correspondence in classically chaotic systems.

\acknowledgements

We thank Prof. C.-H. Lai for his kind support and encouragement.
J.W. acknowledges financial support from Defence Science and
Technology Agency (DSTA) of Singapore under agreement of
POD0613356. J.G. is supported by the start-up fund (WBS grant No.
R-144-050-193-101 and No. R-144-050-193-133) and the NUS ``YIA''
fund (WBS grant No. R-144-000-195-123), both from the National
University of Singapore.

\end{document}